\documentclass[twocolumn,secnumarabic,amssymb,aps,prb]{revtex4-2}

\usepackage{graphicx}
\usepackage{bm}
\usepackage{color}
\usepackage{ulem}

\begin{document}
\title{Entanglement-based tensor-network strong-disorder renormalization group}

\author{Kouichi Seki$^1$}
\author{Toshiya Hikihara$^2$}
\author{Kouichi Okunishi$^1$}

\affiliation{
  $^1$Department of Physics, Niigata University, Niigata 950-2181, Japan \\
  $^2$Faculty of Science and Technology, Gunma University, Kiryu, Gunma 376-8515, Japan
}

\begin{abstract}
We propose an entanglement-based algorithm of the tensor-network strong-disorder renormalization group (tSDRG) method for quantum spin systems with quenched randomness.
In contrast to the previous tSDRG algorithm based on the energy spectrum of renormalized block Hamiltonians,  we directly utilizes the entanglement structure 
 associated with the blocks to be renormalized.
We examine accuracy of the new algorithm for the random antiferromagnetic Heisenberg models on the one-dimensional, triangular, and square lattices.
We then find that the entanglement-based tSDRG achieves better accuracy than the previous one for the square lattice model with weak randomness, while it is less efficient for the one-dimensional and triangular lattice models particularly in the strong randomness region.
The theoretical background and possible improvements of the algorithm are also discussed.
\end{abstract}

\date{\today}

\maketitle

\section{Introduction}\label{sec:Intro}

The tensor network has been under intensive studies in the fields of condensed-matter physics and quantum information for the last decades.
Several numerical algorithms based on the tensor-network formalism have been developed and widely applied to quantum many-body systems to efficiently extract their low-energy states.
The most successful example is the density-matrix renormalization group (DMRG) method\cite{White1992,White1993}, which can be viewed 
as a variational method based on the matrix-product state\cite{OstlundR1995,RommerO1997} and has achieved extremely accurate calculations for one-dimensional (1D) quantum many-body systems.
The tensor-product state\cite{NishinoHOMAG2001} and the projected entangled-pair state\cite{VerstraeteC2004,VerstraeteWPC2006} also provide powerful numerical algorithms in exploring quantum many-body systems in two or higher dimensions.
The multiscale entanglement renormalization ansatz\cite{Vidal2007,EvenblyV2009} has also succeeded in efficiently describing the quantum criticality in one dimension.

\begin{figure}
\begin{center}
\includegraphics[width=70mm]{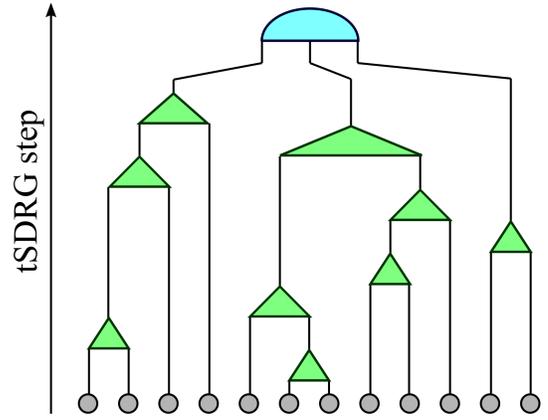}
\caption{
Schematic picture of the tree-tensor network constructed in the tSDRG.
Circles on the bottom and triangles in the middle layers respectively represent the original spins and  the renormalization matrices $V$ or $U$.
The semicircle at the top represents the ground-state wave function in the truncated basis at the final step of tSDRG.
See the text in Sec.\ \ref{sec:Algorithm} for the details.
}
\label{fig:TTN}
\end{center}
\end{figure}

The focus of this work is on how to develop a tensor-network approach to quantum spin systems with quenched randomness.
In 1D random quantum systems, the interplay of quantum fluctuation and randomness often leads to such an exotic state as random-singlet state\cite{Fisher1994}.
It has been also revealed by recent studies\cite{WatanabeKNS2014,ShimokawaWK2015,KawamuraWS2014,UematsuK2017,UematsuK2018,LiuSLGS2018,KimchiNS2018,KimchiSML2018,WuGS2019,KawamuraU2019,UematsuK2019,RenXWSG2020,UematsuHK2020} that when random spin systems contain strongly frustrating interactions, a novel state, sometimes called ``frustrated random-singlet state", may emerge not only in 1D but also higher-dimensional systems.
However, practical numerical studies of the exotic states in frustrated random systems are limited in the level of small size clusters so far;
The quantum Monte-Carlo (QMC) simulation is basically not applicable to frustrated systems due to the minus-sign problem.
The DMRG method is less efficient for higher-dimensional systems and often suffers from quasi-degenerate ground states in random systems\cite{UematsuHK2020}.
Then, a promising numerical approach is the tensor-network strong-disorder renormalization group (tSDRG) method\cite{HikiharaFS1999,GoldsbroughR2014}.
The tSDRG was introduced as an extension of the perturbative strong-disorder renormalization group (SDRG) method\cite{MaDH1979,DasguptaM1980} and has proven to be efficient for realizing accurate numerical calculations of 1D random quantum spin systems\cite{HikiharaFS1999,GoldsbroughR2014,LinKCL2017,TsaiCL2020}.
In the context of tensor network, the tSDRG is based on the tree-tensor network (TTN), as depicted in Fig.\ \ref{fig:TTN};
As tSDRG iterations proceed, the spins in the system are renormalized to form blocks of spins from bottom to top, eventually yielding the TTN representing the ground-state wavefunction of the whole system.

In the tSDRG previously developed, the Hamiltonian is renormalized into the effective one expressed in the truncated basis by the renormalization group (RG) transformation based on the low-energy spectrum and the corresponding eigenstates of intra- and inter-block Hamiltonians.
We thus obtain the low-energy effective Hamiltonian that gives an approximated ground state.
The idea of this {\it Hamiltonian-based} tSDRG (H-tSDRG) 
is faithful to the strategy of the conventional real-space RG, 
where successive RG transformations result in the fixed-point Hamiltonian.
From the perspective of the variational method based on the TTN, nevertheless, there can be an alternative strategy;
For generating an optimal TTN approximating the true ground state, one may construct the RG transformation based not on the low-energy spectrum of block Hamiltonians but on the entanglement structure between blocks in the ground state of the whole system.
The aim of this study is to develop this {\it entanglement-based} tSDRG (E-tSDRG) algorithm.

In this paper, we develop a new algorithm of the E-tSDRG and then apply 
it to the $S=1/2$ antiferromagnetic (AF) Heisenberg models with random exchange couplings defined on 1D chain, triangular, and square lattices.
The model Hamiltonian is formally written as 
\begin{eqnarray}
\mathcal{H}=\sum_{i,j} J_{i,j} {\bm S}_i \cdot {\bm S}_j,
\label{eq:Ham}
\end{eqnarray}
where ${\bm S}_i$ is the spin-1/2 operator at the $i$th site.
Here, the exchange constant $J_{i,j}$ for the nearest-neighboring pairs $(i,j)$ takes a nonzero random value obeying the uniform distribution between $[ 1-\delta, 1+\delta]$, 
\begin{eqnarray}
P(J_{i,j}) = \frac{1}{2\delta} \Theta(J_{i,j}-1+\delta) \Theta(1+\delta-J_{i,j}),
\label{eq:exJ_distribution}
\end{eqnarray}
where $0 < \delta \le 1$ and $\Theta(x)$ is the Heaviside step function.
For the pairs $(i,j)$ that are not on the nearest neighbor, we set $J_{i,j}=0$.
We apply the E-tSDRG algorithm to finite-size clusters of the model (\ref{eq:Ham}) and then compare its accuracy with that of the H-tSDRG.
We thereby show that the E-tSDRG can be more precise than the H-tSDRG for the square lattice model with small randomness, while the E-tSDRG is basically accurate but less efficient for the 1D and triangular-lattice models.

The rest of the paper is organized as follows.
In Sec.\ \ref{subsec:H-tSDRG} a brief review of the  H-tSDRG algorithm is presented, while in Sec.\ \ref{subsec:tSDRG-EE} the new E-tSDRG algorithm is introduced.
The numerical results are presented in Sec.\ \ref{sec:NumRes}.
Section\ \ref{sec:Conc} is devoted to the summary and discussions.

\section{Algorithm}\label{sec:Algorithm}
\subsection{H-tSDRG algorithm}\label{subsec:H-tSDRG}

\begin{figure}
\begin{center}
\includegraphics[width=75mm]{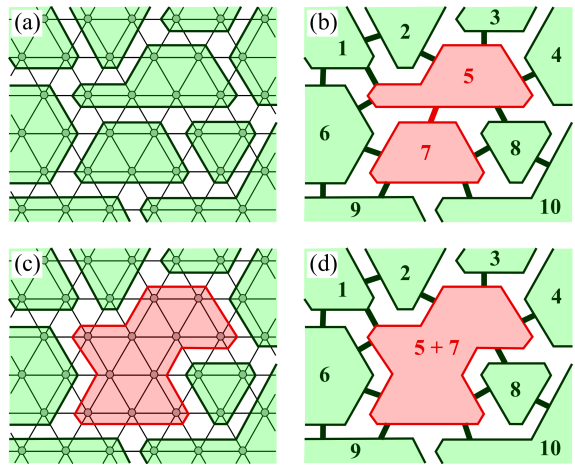}
\caption{
(a) The system after a certain number of H-tSDRG iterations and (b) the corresponding block representation of the system.
Here, we suppose that the link between the blocks $(R,R')=(5,7)$ is the ``strongest" (i.e., has the largest gap $\Delta$) so that the (5,7) blocks are renormalized in this H-tSDRG step.
The system after the renormalization process is depicted in (c) and (d).
}
\label{fig:RGstep_prev}
\end{center}
\end{figure}

Let us begin with a brief review of the H-tSDRG algorithm based on the energy spectrum of the block Hamiltonians, before proceeding to 
details of the E-tSDRG algorithm.
In the tSDRG, the system is generally treated as an assembly of blocks consisting of original spins. (See Fig.\ \ref{fig:RGstep_prev}.)
The Hamiltonian of the whole system is written as
\begin{eqnarray}
\mathcal{H} = \sum_r \mathcal{H}^{\rm B}_r + \sum_{r,r'} \mathcal{H}^{\rm I}_{r,r'},
\label{eq:Ham_tSDRG}
\end{eqnarray}
where 
\begin{eqnarray}
\mathcal{H}^{\rm B}_r = \sum_{i,j \in r} J_{i,j} {\bf S}_i \cdot {\bf S}_j
\label{eq:Ham_B}
\end{eqnarray}
is the intrablock Hamiltonian for $r$th block and 
\begin{eqnarray}
\mathcal{H}^{\rm I}_{r,r'} = \sum_{i\in r, j\in r'} J_{i,j} {\bf S}_i \cdot {\bf S}_j
\label{eq:Ham_I}
\end{eqnarray}
is the interblock Hamiltonian between $r$th and $r'$th blocks.
We assume that the Hilbert space of each block is spanned by $\chi$-dimensional bases, after certain renormalization iterations.
Therefore, the intrablock Hamiltonian $\mathcal{H}^{\rm B}_r$ and the original spin operator $S^\alpha_i$ ($\alpha=x,y,z$) are represented by $\chi \times \chi$ matrices, while the interblock Hamiltonian $\mathcal{H}^{\rm I}_{r,r'}$ is a $\chi^2 \times \chi^2$ matrix \cite{degen_eigenstates}.

The H-tSDRG 
algorithm consists of the following two processes.
The first process is to identify the block pair that is connected by the ``strongest" link.
The second process is to renormalize the block pair into a new single block represented with truncated bases.
In the first process, the strength of link is evaluated by an appropriate energy gap $\Delta_{r,r'}$ in the energy spectrum of the interblock Hamiltonian $\mathcal{H}^{\rm I}_{r,r'}$ or the block-pair Hamiltonian defined as
\begin{eqnarray}
\mathcal{H}^{\rm P}_{r,r'} = \mathcal{H}^{\rm B}_r + \mathcal{H}^{\rm B}_{r'} + \mathcal{H}^{\rm I}_{r,r'}.
\label{eq:Ham_pair}
\end{eqnarray}
Then, the strongest link is determined as the link with the largest $\Delta_{r,r'}$ among all the links of the nonzero interblock Hamiltonians.
Here, note that there are several options for the definition of the energy gap $\Delta_{r,r'}$, which affect resulting accuracy.
In Sec.\ \ref{sec:NumRes}, we adopt two types of energy gaps, $\Delta^{\rm I}_{\rm max}$ and $\Delta^{\rm P}_{\rm gs}$,
which respectively denote the largest level spacing in the energy spectrum of $\mathcal{H}^{\rm I}_{r,r'}$ and the gap between the ground-state and first-excited multiplets of $\mathcal{H}^{\rm P}_{r,r'}$.
It has turned out that the former and the latter respectively yield accurate results for strong- and weak-randomness regimes.
We refer the readers to Ref.\ [\onlinecite{SekiHO2020}] for the detailed definition of $\Delta^{\rm I}_{\rm max}$ and $\Delta^{\rm P}_{\rm gs}$.

Let the block pair $(R,R')$ be the one connected by the strongest link.
In the second process, we renormalize the pair into a new single block.
The Hilbert space of the block pair before the renormalization is spanned by $\chi^2$ bases, and we truncate it into the $\chi$-dimensional space spanned by the $\chi$-lowest energy eigenvectors of the pair-block Hamiltonian $\mathcal{H}^{\rm P}_{R,R'}$\cite{degen_eigenstates}.
The block Hamiltonian of the new block $R+R'$ and the spin operators belonging to the new block are thus obtained 
as
\begin{eqnarray}
\tilde{\mathcal{H}}^{\rm B}_{R+R'} &=& V^\dagger \mathcal{H}^{\rm P}_{R,R'} V,
\label{eq:renorm_Han_B} \\
\tilde{S}^\alpha_i &=& V^\dagger \left( S^\alpha_i \otimes I_{R'} \right) V,
\label{eq:renorm_S_i} \\
\tilde{S}^\alpha_j &=& V^\dagger \left( I_R \otimes S^\alpha_j \right) V,
\label{eq:renorm_S_j}
\end{eqnarray}
where $i \in R$, $j \in R'$, and $I_R$ ($I_{R'}$) is the identity matrix for the block $R$ ($R'$).
The renormalization matrix $V$ is composed of the $\chi$-lowest-energy eigenvectors of $\mathcal{H}^{\rm P}_{R,R'}$.
The interblock Hamiltonian between the new block $R+R'$ and the block $R''$ connected to $R+R'$ via a nonzero interblock Hamiltonian is also renormalized as
\begin{eqnarray}
\tilde{\mathcal{H}}^{\rm I}_{R+R', R''} = V^\dagger \left( \mathcal{H}^{\rm I}_{R, R''} \otimes I_{R'} + I_R \otimes \mathcal{H}^{\rm I}_{R',R''} \right) V.
\label{eq:renorm_Ham_I}
\end{eqnarray}

As a result of the above RG processes, the number of blocks in the whole system is reduced by one.
We continue 
H-tSDRG iterations until the system is represented by three blocks, where the Hamiltonian of the whole system can be exactly diagonalized within the truncated basis.
As shown in Fig\ \ref{fig:TTN}, the resulting ground-state wavefunction is represented as a TTN, in which the isometries corresponding to the renormalization matrices are connected via the $\chi$-dimensional bonds.
Using this TTN, one can straightforwardly calculate the ground-state expectation values of observables.
The algorithm of the H-tSDRG is summarized in Tab.\ \ref{tab:tSDRG_prev}.
We note that the H-tSDRG and the E-tSDRG discussed in the following section 
become exact if $\chi$ reaches the dimension of the total Hilbert space of the system with no cutoff.

\begin{table}
\caption{
Algorithm of the H-tSDRG.
}
\label{tab:tSDRG_prev}
\begin{center}
\begin{tabular}{ll}
\hline
1. & Calculate the energy gap $\Delta_{r,r'}$ for all block pairs \\
   & connected by  nonzero interblock Hamiltonians. \\
2. & Identify the block pair $(R,R')$ with the largest $\Delta_{r,r'}$. \\
3. & Diagonalize the block-pair Hamiltonian $\mathcal{H}^{\rm P}_{R,R'}$ to obtain \\
   & the renormalization matrix $V$. \\
4. & Using $V$, renormalize $\mathcal{H}^{\rm P}_{R,R'}$ and $S^\alpha_i$ ($i \in R$ or $R'$) to\\
   & obtain the new-block Hamiltonian $\mathcal{H}^{\rm B}_{R+R'}$ and the spin \\
   & operators $S^\alpha_i$ in the new block $R+R'$. Renormalize also\\
   & the interblock Hamiltonians $\tilde{\mathcal{H}}^{\rm I}_{R+R', R''}$ between the new \\
   & block $R+R'$ and a block $R''$ linked to the new block via  \\
   & nonzero interblock Hamiltonians. \\
5. & Diagonalize the new interblock or block-pair Hamiltonians \\
   & to renew the list of the energy gap $\Delta_{r,r'}$. \\
6. & Back to 2. \\
\hline
\end{tabular}
\end{center}
\end{table}

\subsection{E-tSDRG algorithm}\label{subsec:tSDRG-EE}

As discussed in the preceding section, the H-tSDRG can be formulated as a variational method using the TTN wavefunction, where its accuracy depends on both the network structure and the quality of isometries.
In the TTN, a branch and the rest of the tree are generally connected by a single bond with a finite dimension $\chi$, which can carry an entanglement entropy 
of up to $\ln \chi$.
This fact naturally provides the guiding principle that the block pair having the smallest entanglement with its environment should be renormalized first.

In the H-tSDRG, the block pair connected by the ``strongest" link is renormalized first, referring to the excitation gap in the spectrum of ${\cal H}^{\rm I}$ or ${\cal H}^{\rm P}$.
This is basically consistent with the above guiding principle, 
from the monogamy of entanglement stating that two blocks coupled strongly with each other have a small entanglement with their environment.
However, this scheme based on the energy spectrum is quite indirect to 
see entanglement structures around the block pairs.
Also, the isometries in the H-tSDRG are constructed from the low-energy eigenstates of $\mathcal{H}^{\rm P}_{R,R'}$, where the entanglement effect from the environment around the block pair may not be included sufficiently.
Thus, a tSDRG algorithm based on more direct use of the entanglement distribution around the blocks should be examined particularly for random spin systems.

Let us consider the reduced density matrix (rDM) for a block pair $(r,r')$ in the ground state of the whole system,
\begin{eqnarray}
\rho_{r,r'} = {\rm Tr}_{\overline{r,r'}} | \Psi_{\rm g} \rangle \langle \Psi_{\rm g} |,
\label{eq:subDMrrp}
\end{eqnarray}
where $| \Psi_{\rm g} \rangle$ is the ground state of the whole system and ${\rm Tr}_{\overline{r,r'}}$ denotes the trace 
with respect to the degrees of freedom complemental to the $r$th and $r'$th blocks.
From $\rho_{r,r'}$, one can extract the entanglement entropy between the block pair $(r,r')$ and the rest of the system, which can be used for identifying the block pair $(R,R')$ to be renormalized.
Also, the eigenvectors of $\rho_{R,R'}$ with the $\chi$-largest eigenvalues provide the isometry that takes account of the entanglement between the block pair and the environment.
Of course, it is generally difficult to obtain the exact rDM of Eq.\ (\ref{eq:subDMrrp}) a priori and thus we should introduce an appropriate approximation.

\begin{figure}
\begin{center}
\includegraphics[width=60mm]{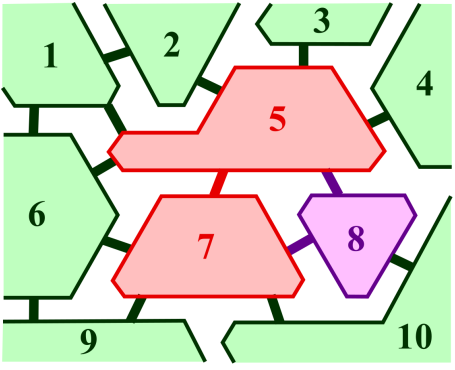}
\caption{
The blocks involved in the rDM $\rho_{r,r'}(r'')$ [Eq.\ (\ref{eq:rDM_three})] with $(r,r')=(5,7)$ and $r''=8$.
The calculation of $\tilde{\rho}_{r,r'}$ [Eq.\ (\ref{eq:rDM_mixed})] for $(r,r')=(5,7)$ requires $\rho_{5,7}(r'')$ with  $r''=1,2,3,4,6,8,9,10$.
}
\label{fig:RGstep_EE}
\end{center}
\end{figure}

Here, we propose the following way of approximating the rDM of Eq. (\ref{eq:subDMrrp}).
Let $(r,r')$ be the indices for the block pair and $r''$ be the index for the blocks linked to the $(r,r')$ blocks via a nonzero interblock Hamiltonian.
(See Fig.\ \ref{fig:RGstep_EE}.)
The three-block Hamiltonian for the blocks $(r,r',r'')$ is given by
\begin{eqnarray}
\mathcal{H}^{(3)}_{r,r',r''} = \mathcal{H}^{\rm P}_{r,r'} + \mathcal{H}^{\rm B}_{r''} + \mathcal{H}^{\rm I}_{r,r''} + \mathcal{H}^{\rm I}_{r',r''}.
\label{eq:Ham_three}
\end{eqnarray}
Using the ground-state wavefunction of the three-block Hamiltonian, $|\varphi_{\rm g}\rangle_{r,r',r''}$, we calculate the rDM for the blocks $(r,r')$ in the three-block ground state as
\begin{eqnarray}
\rho_{r,r'}(r'') = {\rm Tr}_{r''} | \varphi_{\rm g} \rangle_{r,r',r''}~_{r,r',r''}\langle \varphi_{\rm g} |,
\label{eq:rDM_three}
\end{eqnarray}
where the trace ${\rm Tr}_{r''}$ is taken for the Hilbert space of the $r''$th block.
Then, we perform calculations of $\rho_{r,r'}(r'')$ for all blocks $r''$ connected to the $(r,r')$ blocks and obtain their mixed-state rDM,
\begin{eqnarray}
\tilde{\rho}_{r,r'} = \frac{1}{D_{r,r'}} \sum_{r''} \rho_{r,r'}(r''),
\label{eq:rDM_mixed}
\end{eqnarray}
which can be used as an approximation of Eq. (\ref{eq:subDMrrp}).
Here, $D_{r,r'}$ denotes the number of the blocks $r''$ linked to the block pair $(r,r')$ via a nonzero interblock Hamiltonian.
We note that in the mixed-state rDM $\tilde{\rho}_{r,r'}$, the weight for $\rho_{r,r'}(r'')$ is assumed to be equally $1/D_{r,r'}$ for all $r''$.
This is because detailes of the weight are not relevant to $\tilde{\rho}_{r,r'}$ as long as the decay of the eigenvalue spectrum of each $\rho_{r,r'}(r'')$ is not too slow.

Using Eq. (\ref{eq:rDM_mixed}), we evaluate the entanglement entropy between the block pair $(r,r')$ and the rest of the system as
\begin{eqnarray}
\mathcal{S}_{r,r'} = - {\rm Tr}_{r,r'} \tilde{\rho}_{r,r'} \ln \tilde{\rho}_{r,r'}.
\label{eq:EEnt_mixed}
\end{eqnarray}
We calculate Eq.\ (\ref{eq:EEnt_mixed}) for all the pairs of $(r,r')$ linked via nonzero interblock Hamiltonians, and then determine the block pair $(R,R')$ to be renormalized as the one having the minimum $\mathcal{S}_{r,r'}$.
Besides, we employ the eigenvectors corresponding to the $\chi$-largest eigenvalues of the rDM $\tilde{\rho}_{R,R'}$ as the bases $\{ {\bm u}_1, ..., {\bm u}_\chi \}$ for the new renormalized block.

We note an additional treatment for the construction of the new block bases, which is required for maintaining the accuracy of practical computations.
For a precise argument, let us write the rank of the mixed-state rDM $\tilde{\rho}_{R,R'}$ as $n$ below.
We may usually expect that $n \gg \chi$ for the block pair $(R,R')$ whose Hilbert space is much larger than $\chi$.
However, we sometimes encounter the situation where $n < \chi$, despite that the block pair $(R,R')$ has a sufficiently large Hilbert space.
This is because the matrix rank of the rDM $\rho_{r,r'}(r'')$ [Eq.\ (\ref{eq:rDM_three})] is generally bounded not only by the dimension of the Hilbert space of the block pair $(r,r')$ but also by that of the block $r''$.
Consequently, when the block pair $(R,R')$ is surrounded only by a small number of blocks with small Hilbert spaces, the rank $n$ can be smaller than $\chi$, even if the block pair $(R,R')$ has a Hilbert space with a dimension larger than $\chi$.
Note that the dimension of the Hilbert space for the block $r''$ can be smaller than $\chi$ when the block $r''$ contains only a small number of spins.
If this ``small-rank problem" is the case, we can prepare only $n (< \chi)$ bases for the Hilbert space of the new block.
In this case, the entanglement in the mixed-state rDM $\tilde{\rho}_{R,R'}$ [Eq.\ (\ref{eq:rDM_mixed})] can be indeed maintained within the $n$ bases.
However, if one keeps only those $n$ bases, some essential entanglements between the block pair $(R,R')$ and its distant blocks, that are not involved in $\tilde{\rho}_{R,R'}$, may be missed, resulting in loss of computational accuracy.
To fix the problem, we examine the following additional treatment.

\begin{figure}
\begin{center}
\includegraphics[width=75mm]{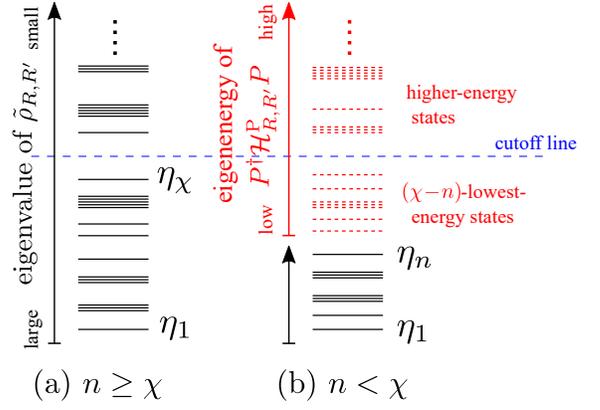}
\caption{
Schematic graph of the new-block bases in the E-tSDRG algorithm;
(a) $n \ge \chi$ (usual case) and (b) $n < \chi$ (the case of the small-rank problem).
Solid lines (black) represent the eigenstates of the mixed-state rDM $\tilde{\rho}_{R,R'}$ and the dotted lines (red) represent the eigenstates of the block-pair Hamiltonian in the projected space, $P^\dagger \mathcal{H}^{\rm P}_{R,R'} P$.
Dashed line (blue) denotes the cutoff line.
}
\label{fig:add_treat}
\end{center}
\end{figure}

Let $\{ {\bm \eta}_1, \cdots, {\bm \eta}_\xi \}$ be the orthonormal eigenvectors of $\tilde{\rho}_{R,R'}$.
Here, $\xi$ is the dimension of the Hilbert space for the block pair $(R, R')$.
We then assume that $\{ {\bm \eta}_1, \cdots, {\bm \eta}_n \}$ belong to the $n$ nonzero eigenvalues of $\tilde{\rho}_{R,R'}$, while $\{ {\bm \eta}_{n+1}, \cdots, {\bm \eta}_\xi \}$ corresponds to the eigenvectors of $(\xi-n)$ number of the zero eigenvalue.
We first adopt the eigenvectors $\{ {\bm \eta}_1, \cdots, {\bm \eta}_n \}$ as the $n$ bases of the new block, $\{ {\bm u}_1, \cdots, {\bm u}_n \}$.
Next, using the projection matrix $P$ constructed from $\{ {\bm \eta}_{n+1}, \cdots, {\bm \eta}_\xi \}$, we calculate the block-pair Hamiltonian $P^\dagger \mathcal{H}^{\rm P}_{R,R'} P$ restricted in the reduced Hilbert space orthogonal to $\{ {\bm u}_1, \cdots, {\bm u}_n \}(=\{ {\bm \eta}_1, \cdots, {\bm \eta}_n \} )$.
Diagonalizing $P^\dagger \mathcal{H}^{\rm P}_{R,R'} P$, we then generate the $(\chi-n)$ eigenvectors with the $(\chi-n)$-lowest eigenenergies.
We adopt these eigenvectors as the supplementary bases $\{ {\bm u}_{n+1}, \cdots, {\bm u}_\chi \}$ to obtain the new block bases $\{ {\bm u}_1, \cdots, {\bm u}_\chi \}$.
(See Fig.\ \ref{fig:add_treat}.)
Using the reinforced bases, we can construct the Hilbert space of the new block without accidental reduction of its dimension.
Of course, this is an ad-hoc approach and the augmented bases of the $(\chi-n)$-lowest-energy eigenstates of $P^\dagger \mathcal{H}^{\rm P}_{R,R'} P$ may not be the optimal ones to retain the entanglement between the block pair $(R,R')$ and its distant blocks.
However, we will see in Sec.\ \ref{sec:NumRes} that this procedure actually improves the accuracy of the algorithm.

Now we present the algorithm of  E-tSDRG as follows.
Suppose that, at a certain iteration step of E-tSDRG, the entanglement entropy $\mathcal{S}_{r,r'}$ [Eq.\ (\ref{eq:EEnt_mixed})] 
is obtained for all the pairs of $(r,r')$ linked via a nonzero interblock Hamiltonian $\mathcal{H}^{\rm I}_{r,r'}$.
Using the list of the entanglement entropies, we select the block pair $(R,R')$ having the smallest entanglement entropy as the pair to be renormalized.
For the bases of the new block, we employ the eigenvectors of the $\chi$-largest eigenvalues of the rDM $\tilde{\rho}_{R,R'}$\cite{degen_eigenstates}.
If the small-rank problem occurs, the additional treatment above is invoked.
The intrablock Hamiltonian and spin operators involved in the new block as well as the interblock Hamiltonians including the new block are then renormalized as 
\begin{eqnarray}
\tilde{\mathcal{H}}^{\rm B}_{R+R'} &=& U^\dagger \mathcal{H}^{\rm P}_{R,R'} U,
\label{eq:renorm_Han_B_ent} \\
\tilde{S}^\alpha_i &=& U^\dagger \left( S^\alpha_i \otimes I_{R'} \right) U,
\label{eq:renorm_S_i_ent} \\
\tilde{S}^\alpha_j &=& U^\dagger \left( I_R \otimes S^\alpha_j \right) U,
\label{eq:renorm_S_j_ent} \\
\tilde{\mathcal{H}}^{\rm I}_{R+R', R''} &=& U^\dagger \left( \mathcal{H}^{\rm I}_{R, R''} \otimes I_{R'} + I_R \otimes \mathcal{H}^{\rm I}_{R',R''} \right) U,
\label{eq:renorm_Ham_I}
\end{eqnarray}
where $U$ is the (complemented) RG transformation matrix composed of the basis vectors $\{ {\bm u}_1, \cdots, {\bm u}_\chi \}$.
Finally, we calculate the rDM $\tilde{\rho}_{R+R',R''}$ for all the block pairs consisting of the new block $R+R'$ and a block $R''$ linked to the new block, and then, update the list of the entanglement entropies.
We iterate the calculation until the number of blocks becomes small enough so that the Hamiltonian of the whole system in the truncated bases is diagonalizable.
Using the TTN constructed from the set of $U$, we can calculate the expectation values of observables.
The algorithm is summarized in Tab.\ \ref{tab:tSDRG_ent}.

\begin{table}
\caption{
Algorithm of the E-tSDRG.
}
\label{tab:tSDRG_ent}
\begin{center}
\begin{tabular}{ll}
\hline
1. & Calculate the entanglement entropy $\mathcal{S}_{r,r'}$ between a \\
   &block pair $(r,r')$ and its environment for all block pairs\\
   & connected via the nonzero interblock Hamiltonian $\mathcal{H}^{\rm I}_{r,r'}$. \\
2. & Identify the block pair $(R,R')$ with the smallest \\
   & entanglement entropy $\mathcal{S}_{r,r'}$. \\
3. & Calculate and diagonalize the rDM $\tilde{\rho}_{R,R'}$ [Eq.\ (\ref{eq:rDM_mixed})] to \\
   & obtain the renormalization matrix $U$. (Adopt the \\
   & additional treatment for the small-rank problem, if \\
   & necessary.) \\
4. & Using $U$, renormalize $\mathcal{H}^{\rm P}_{R,R'}$ and $S^\alpha_i$ ($i \in R$ or $R'$) to\\
   & obtain the intrablock Hamiltonian $\mathcal{H}^{\rm B}_{R+R'}$ and the spin\\
   & operators $S^\alpha_i$ in the new block $R+R'$. Renormalize also\\
   & the interblock Hamiltonians $\tilde{\mathcal{H}}^{\rm I}_{R+R', R''}$ between the new \\
   & block $R+R'$ and a block $R''$ linked to the new block via  \\
   & nonzero interblock Hamiltonians. \\
5. & Calculate the rDM and entanglement entropy for the \\
   & block pairs containing the new block and update \\
   & the list of the entanglement entropy $\mathcal{S}_{r,r'}$. \\
6. & Back to 2. \\
\hline
\end{tabular}
\end{center}
\end{table}

Before closing this section, we mention the computational cost of the E-tSDRG.
The most costly part in the E-tSDRG algorithm is the process to update the list of the entanglement entropy after the renormalization.
The process requires calculations of the mixed-state rDM $\tilde{\rho}_{R+R',R''}$ for all the block pairs $(R+R',R'')$ composed of the new renormalized block $R+R'$ and a block $R''$ linked to it.
Thus, its total computational cost is proportional to the construction and diagonalization of $\rho_{R+R',R''}(r'')$ for three-block clusters $(R+R',R'',r'')$ with two running indices $R''$ and $r''$ [see Eq.\ (\ref{eq:rDM_mixed})].
Note that the process in the H-tSDRG corresponding to this part is the update of the list of the energy gap $\Delta_{R+R',R''}$ including only a single running index $R''$.
Accordingly, the E-tSDRG calculation in each iteration step is heavier than that of the H-tSDRG by about a factor of $D$, the average of the block coordination number $D_{r,r'}$.
Meanwhile, in both the E- and H-tSDRGs, the maximum memory array is equally bounded by the full diagonalization of $\chi^2 \times \chi^2$ matrices, $\tilde{\rho}_{r,r'}$, $\mathcal{H}^{\rm I}_{r,r'}$, or $\mathcal{H}^{\rm P}_{r,r'}$.
Therefore, the maximum value of $\chi$ taken in a practical calculation would be more or less the same in the E- and H-tSDRGs.

\section{Numerical results}\label{sec:NumRes}

In this section, we discuss the numerical accuracy of the E-tSDRG algorithm applied to the random AF Heisenberg models defined on the 1D chain, the triangular lattice, and the square lattice.
We particularly compare benchmark results of the E-tSDRG for finite-size clusters with 
those of the H-tSDRG.
More precisely, we treat the 1D chains with $N=24$ spins, the triangular-lattice systems with  $N=24$ and $36$ spins (the shape of the $N=24$ cluster is as in Ref.\ [\onlinecite{WuGS2019}] and that of $N=36$ is $6 \times 6$), and the square-lattice systems with $N=36$ ($6 \times 6$) spins.
The periodic boundary conditions are imposed in the all cases.
The number of random samples $\mathcal{N}_{\rm s}$ is $\mathcal{N}_{\rm s}=1000$ for the 1D chain and $\mathcal{N}_{\rm s}=500$ for the triangular and square lattices.
We have confirmed that the error due to the random sampling is small enough for the following arguments.
The bond dimension used in 
tSDRG calculations is up to $\chi=80$.
For the above-mentioned finite-size clusters except for the triangular-lattice system with $N=36$, we examine the deviation of the E-tSDRG and H-tSDRG results from the exact (or pseudo-exact) ones obtained by the exact diagonalization (ED) and QMC simulation\cite{SyljuaasenS2002,KawashimaH2004}.
For the $N=36$ triangular-lattice cluster, the exact data are not available and thus we directly compare the E-tSDRG and H-tSDRG results.
The numerical data of the ED, QMC, and H-tSDRG calculations are equivalent to those presented in Ref.\ [\onlinecite{SekiHO2020}]\cite{samplenumber}.
We note that the QMC simulation was done for a 
temperature low enough to describe the ground state\cite{QMC-SPEC}.
In addition, the data of the H-tSDRG 
 were obtained by the algorithm with $\Delta^{\rm I}_{\rm max}$ and $\Delta^{\rm P}_{\rm gs}$, which provided accurate results respectively in strong and weak randomness regimes\cite{SekiHO2020}.

\begin{figure}
\begin{center}
\includegraphics[width=70mm]{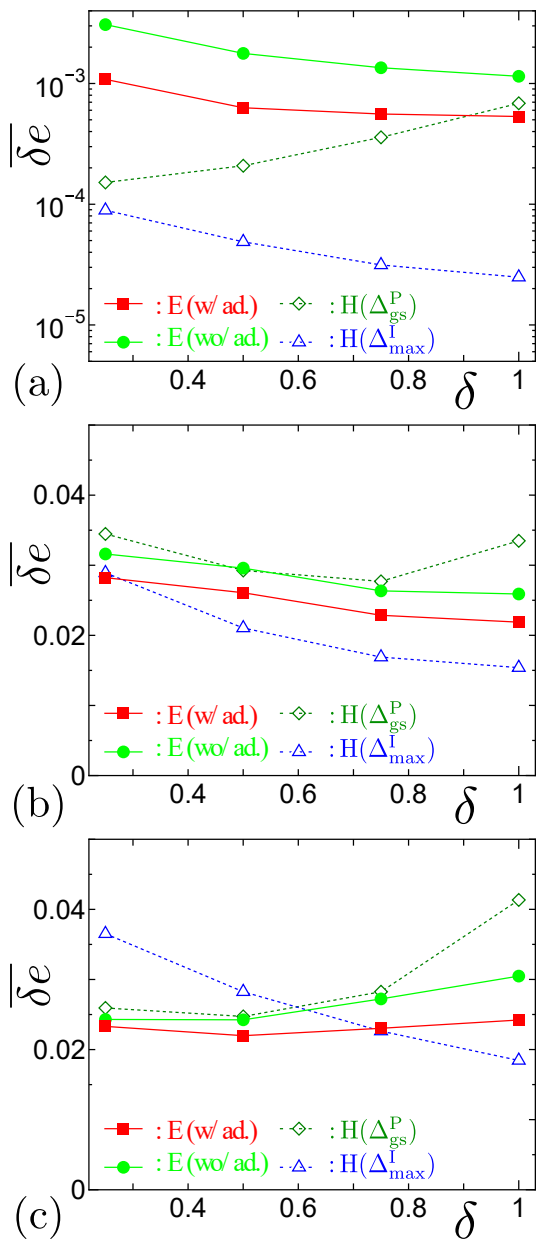}
\caption{
Random averages of the errors of the ground-state energy per spin, $\overline{\delta e}$, as functions of $\delta$ for (a) 1D chain with $N=24$ spins, (b) triangular lattice with $N=24$ spins, and (c) square lattice with $N=36$ spins.
Solid squares and circles represent respectively the results of E-tSDRG with and without the additional treatment for the small-rank problem (see text in Sec.\ \ref{subsec:tSDRG-EE}).
Open triangles and diamonds show the results of the H-tSDRG using $\Delta^{\rm I}_{\rm max}$ and $\Delta^{\rm P}_{\rm gs}$, respectively.
The bond dimension used in the calculations is $\chi=80$.
The data in (a) are plotted in a logarithmic scale while the data in (b) and (c) are in a linear scale.
}
\label{fig:Ene_error-Dlt}
\end{center}
\end{figure}

Let us start with the random average of the errors of the ground-state energy per spin defined by
\begin{eqnarray}
\overline{\delta e} \equiv \frac{1}{\mathcal{N}_{\rm s} N} \sum_{\nu=1}^{\mathcal{N}_{\rm s}} \left( E_\nu - E_{\nu}^{\rm ex} \right),
\label{eq:Ene_error}
\end{eqnarray}
where $E_\nu$ and $E_{\nu}^{\rm ex}$ indicate respectively the tSDRG and exact results of the ground-state energy of the $\nu$th sample.
%
Figure\ \ref{fig:Ene_error-Dlt} shows $\delta$-dependences of $\overline{\delta e}$ at $\chi=80$.
In the figure, it is verified that the accuracy of the E-tSDRG results is actually improved by the additional treatment for the small-rank problem of the rDM described in Sec.\ \ref{subsec:tSDRG-EE}, suggesting that the small-rank problem has a certain relevance to the numerical accuracy of the E-tSDRG.
It is also found that the improvement by the additional treatment in the calculation of the 1D chain is more significant than that of the two-dimensional lattices.
[Note that the data in Fig.\ \ref{fig:Ene_error-Dlt}(a) are presented in a logarithmic scale.]
Indeed, we have 
confirmed that for the 1D chain, the reduction of the number of the bases 
due to the small-rank problem
manifests itself in the middle and late stages of the E-tSDRG iterations.
This may be because in the 1D chain, the block coordination number $D_{r,r'}$ is always two and the possibility that the mixed-state rDM $\tilde{\rho}_{r,r'}$ has a small rank is relatively high.
For the two-dimensional lattices, on the other hand, the small-rank problem of the rDM is prominent only in the very late stage of the E-tSDRG, where the number of remaining blocks is small and $D_{r,r'}$ becomes also small.

Next, 
we discuss the comparison between the results of the E-tSDRG (with the additional treatment) and those of the H-tSDRG.
It is found in Fig. \ref{fig:Ene_error-Dlt} (a) that for the 1D chain, the H-tSDRG with $\Delta^{\rm I}_{\rm max}$ 
is much more accurate than the E-tSDRG, although the E-tSDRG achieves the accuracy of the order of three digits.
From Figs.\ \ref{fig:Ene_error-Dlt} (b) and (c), on the other hand, it is basically concluded that for the two-dimensional lattices, the E-tSDRG achieves the same order of accuracy as the H-tSDRG.
In particular, the E-tSDRG for the square lattice turns out to be the best in the small $\delta$ region.
The reason why such a qualitative difference of the E-tSDRG occurs depending on the systems is associated with the nature of the approximated rDM $\tilde{\rho}_{r,r'}$, 
where we take account of the entanglements between a target block pair and its environment up to the neighboring blocks.
This implies that the short-range entanglements from the adjacent blocks are overestimated in the rDM, while the entanglements between distant blocks are omitted.
As a result, the rDM fails 
in capturing the effect of distant singlet pairs embedded in the random-singlet state realized in the 1D chain, 
which results in the relatively poor accuracy.
For the square and triangular lattices, meanwhile, the block pairs are usually surrounded by several blocks, 
and thus, the entanglements from the adjacent blocks are relatively significant.
This supports that the E-tSDRG becomes efficient in the small $\delta$ regime of the triangular and square lattices, where the short-range entanglement due to the magnetic ordering has a certain relevance.

\begin{figure}
\begin{center}
\includegraphics[width=70mm]{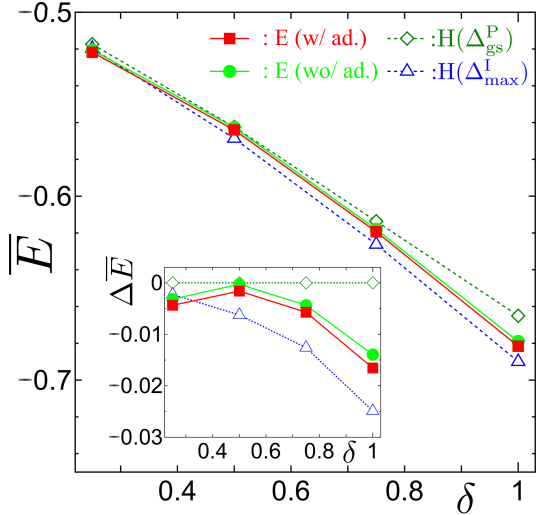}
\caption{
Random averaged ground-state energy per spin, $\overline{E}$, as functions of $\delta$ for the triangular lattice with $N=36$ spins.
Solid squares and circles represent respectively the results of E-tSDRG with and without the additional treatment for the small-rank problem (see text in Sec.\ \ref{subsec:tSDRG-EE}).
Open triangles and diamonds show the results of the H-tSDRG using $\Delta^{\rm I}_{\rm max}$ and $\Delta^{\rm P}_{\rm gs}$, respectively.
The bond dimension used in the calculations is $\chi=80$.
Inset shows the energy gain compared to the data of the H-tSDRG with $\Delta^{\rm P}_{\rm gs}$, $\Delta \overline{E} = \overline{E} - \overline{E}_{\Delta^{\rm P}_{\rm gs}}$. 
}
\label{fig:Ene_av-tri36}
\end{center}
\end{figure}

In order to examine the E-tSDRG for the system sizes beyond the ED level,  we further calculate the random averaged ground-state energy per spin defined as
\begin{eqnarray}
\overline{E} \equiv \frac{1}{\mathcal{N}_{\rm s} N} \sum_{\nu=1}^{\mathcal{N}_{\rm s}} E_\nu ,
\label{eq:Ene_average}
\end{eqnarray}
for the triangular lattice with $N=36$ ($6 \times 6$) spins.
Figure \ \ref{fig:Ene_av-tri36} shows the comparison of the E-tSDRG and H-tSDRG results.
We find that the data exhibit essentially the same tendency as that for the $N=24$ triangular lattice;
The accuracy of the E-tSDRG in the scale of Fig.\ \ref{fig:Ene_av-tri36} is comparable to the H-tSDRG. 
Moreover, the precise comparison in the inset of Fig.\ \ref{fig:Ene_av-tri36} illustrates that the E-tSDRG provides slightly lower energy for small $\delta$, while  the H-tSDRG algorithm with $\Delta^{\rm I}_{\rm max}$ is better for large $\delta$.
This suggests that the E-tSDRG can also treat the larger system size beyond ED, complementarily to the H-tSDRG.

\begin{figure}
\begin{center}
\includegraphics[width=70mm]{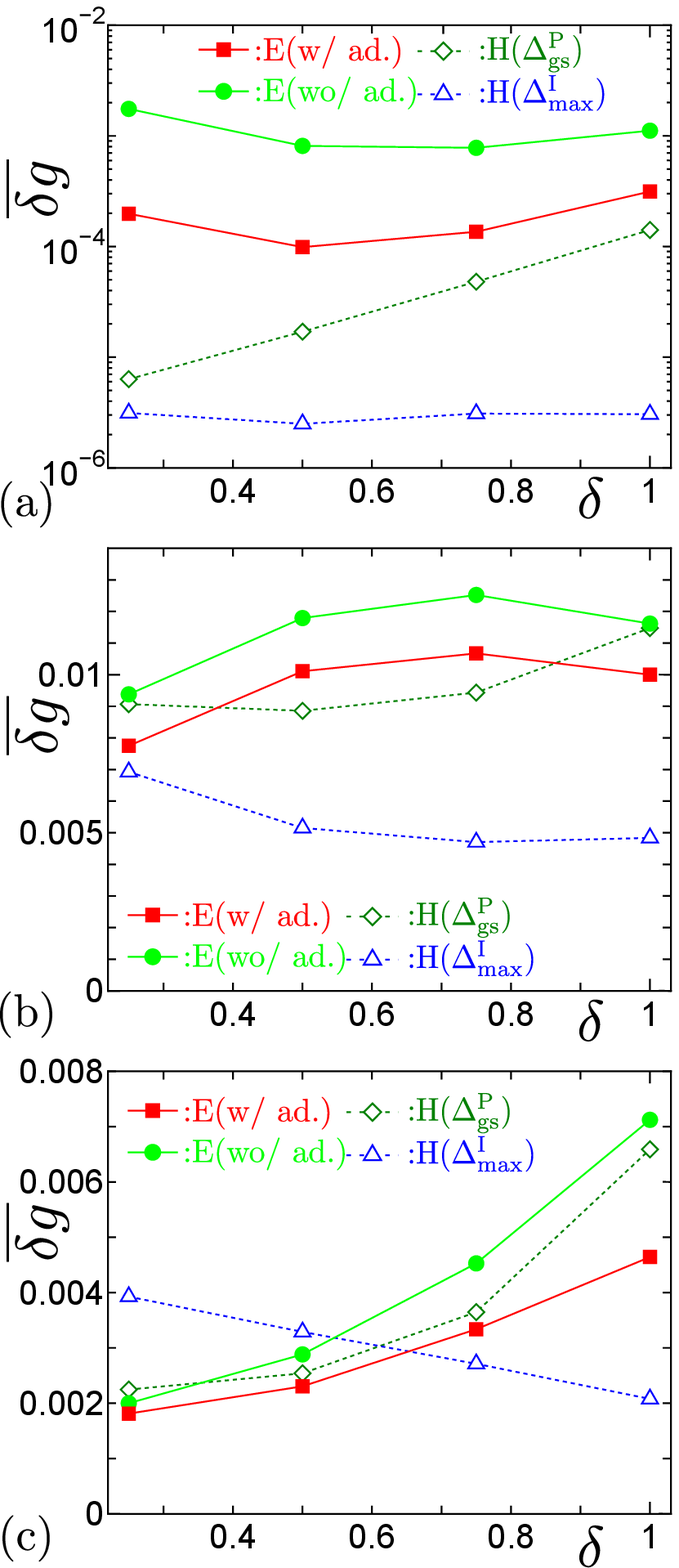}
\caption{
Random averages of the errors of the ground-state correlation functions, $\overline{\delta g}$, as functions of $\delta$ for (a) 1D chain with $N=24$ spins, (b) triangular lattice with $N=24$ spins, and (c) square lattice with $N=36$ spins.
Solid squares and circles represent respectively the results of E-tSDRG with and without the additional treatment for the small-rank problem (see text in Sec.\ \ref{subsec:tSDRG-EE}).
Open triangles and diamonds show the results of the H-tSDRG using $\Delta^{\rm I}_{\rm max}$ and $\Delta^{\rm P}_{\rm gs}$, respectively.
The bond dimension used in the calculations is $\chi=80$.
The data in (a) are plotted in a logarithmic scale while the data in (b) and (c) are in a linear scale.
}
\label{fig:Cor-error-Dlt}
\end{center}
\end{figure}

We also explore the random average of the errors of the ground-state correlation functions,
\begin{eqnarray}
\overline{\delta g} \equiv \frac{1}{\mathcal{N}_{\rm s}} \sum_{\nu=1}^{\mathcal{N}_{\rm s}} \sqrt{\frac{2}{N(N-1)}\sum_{i}\sum_{j(\ne i)} \left[ g_\nu(i,j)-g_{\nu}^{\rm ex}(i,j)\right]^2},
\nonumber \\
\label{eq:cor_error}
\end{eqnarray}
with
$g_\nu(i,j) \equiv \langle {\bm S}_i \cdot {\bm S}_j \rangle_\nu$
and
$g_{\nu}^{\rm ex}(i,j) \equiv \langle {\bm S}_i \cdot {\bm S}_j \rangle_{\nu}^{\rm ex}$.
Here, $\langle \cdots \rangle_\nu$ and $\langle \cdots \rangle_{\nu}^{\rm ex}$ represent the ground-state expectation values respectively calculated by the tSDRGs and exact (ED or QMC) methods.
The $\delta$-dependence of $\overline{\delta g}$ with $\chi=80$ is shown in Fig.\ \ref{fig:Cor-error-Dlt}.
The data indicate that $\overline{\delta g}$ also exhibits the same tendency as that of $\overline{\delta e}$;
For the 1D chain, the accuracy of the E-tSDRG is of the order of four digits, although the H-tSDRG with $\Delta^{\rm I}_{\rm max}$ achieves much better accuracy than the E-tSDRG.
For the two-dimensional lattices, however, the accuracy of E-tSDRG turns out to be comparable to that of H-tSDRGs.
In particular, the E-tSDRG becomes the best for the square lattice in the small $\delta$ region.

\begin{figure}
\begin{center}
\includegraphics[width=70mm]{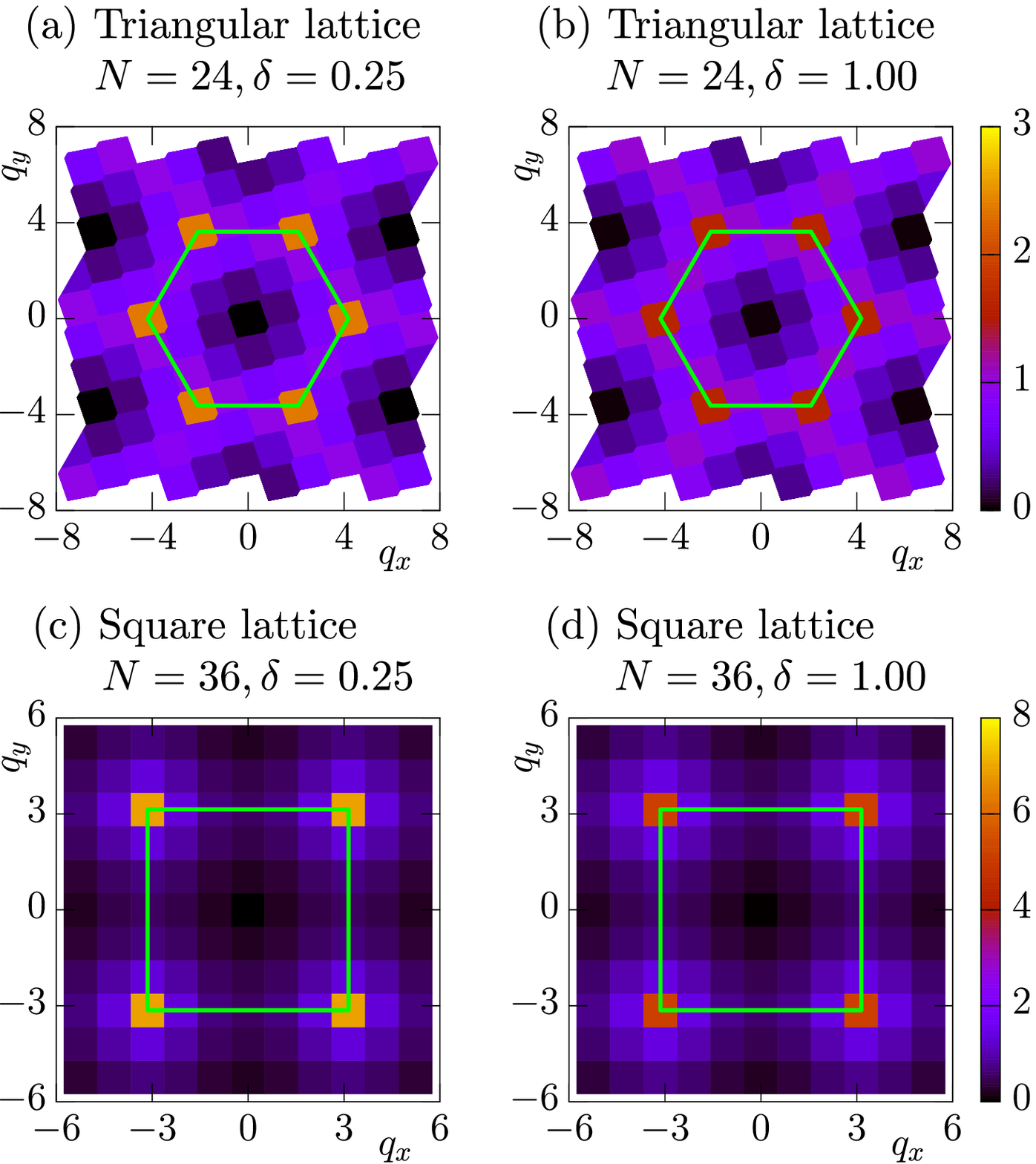}
\caption{
Random-averaged static spin structure factors obtained by the E-tSDRG (with the additional treatment) with $\chi=80$ for (a) triangular lattice with $N=24$ and $\delta=0.25$, (b) triangular lattice with $N=24$ and $\delta=1.00$, (c) square lattice with $N=36$ and $\delta=0.25$, and (d) square lattice with $N=36$ and $\delta=1.00$.
}
\label{fig:Sq}
\end{center}
\end{figure}

In order to confirm if the E-tSDRG captures qualitative features of the systems correctly, we 
compute the random-averaged static spin structure factor,
\begin{eqnarray}
\overline{S}({\bm q}) 
&\equiv& \frac{1}{\mathcal{N}_{\rm s}} \sum_{\nu=1}^{\mathcal{N}_{\rm s}}
\left\langle \left| \frac{1}{\sqrt{N}} \sum_j {\bm S}_j e^{{\rm i}{\bm q}\cdot{\bm r}_j} \right|^2\right\rangle_\nu
\nonumber \\
&=& \frac{1}{\mathcal{N}_{\rm s}N} \sum_{\nu=1}^{\mathcal{N}_{\rm s}}
\sum_{i,j} g_\nu(i,j) \cos\left[ {\bm q}\cdot({\bm r}_i - {\bm r}_j)\right],
\label{eq:Sq}
\end{eqnarray}
where ${\bm r}_j$ is the position vector of the $j$th spin, for the triangular and square lattices.
Figure\ \ref{fig:Sq} shows the results for those lattices with $\delta=0.25$ and $1.00$.
In the all cases, the results correctly reproduce qualitative features such as broad peaks at K points for the triangular-lattice case and 
sharp peaks at ${\bm q}=(\pi, \pi)$ for the square-lattice case.

\section{Concluding remarks}\label{sec:Conc}

In summary, we have proposed an entanglement-based tSDRG (E-tSDRG) algorithm for the random AF Heisenberg models, in comparison with the previous version of tSDRG (H-tSDRG) algorithms based on the energy spectrum of block Hamiltonians.
A key point is that we have directly evaluated the rDM [Eqs.\ (\ref{eq:rDM_three}) and (\ref{eq:rDM_mixed})] of the block pair to be renormalized, by taking account of up to the blocks directly linked to the block pair.
On the basis of the entanglement entropy for the rDM, we then construct the E-tSDRG algorithm 
generating a TTN with a slightly different structure than the H-tSDRG.
In order to evaluate the accuracy of the E-tSDRG, we have applied the method to finite-size clusters of the random AF Heisenberg model up to $N=36$ sites.
We have then demonstrated that the E-tSDRG achieves better accuracy than the H-tSDRG for the square lattice model in the small randomness region, while it is pretty good but less efficient for the 1D chain and the triangular lattice models.

The tendency of the accuracy of the E-tSDRG can be understood also from the viewpoint of the TTN. 
The E-tSDRG is an one-way algorithm constructing the TTN in Fig. \ref{fig:TTN} from the bottom to top by looking ahead the upper branch in the TTN through the rDM of the block pairs.
In the present construction of the rDM, then, the entanglement among distant blocks is missed and thus the short-range correlation is overestimated.
Actually, the E-tSDRG is relatively poor for the random-singlet state of the 1D chain where singlet pairs between distant spins are important\cite{Fisher1994}.
This is also the case for the model (\ref{eq:Ham}) with large $\delta$ in the triangular lattice, 
where the frustrated random-singlet ground state containing a certain amount of long-distance singlet pairs is expected to emerge\cite{WatanabeKNS2014,ShimokawaWK2015,KawamuraU2019}.
Meanwhile, the E-tSDRG can achieve good accuracy for the square-lattice model where the short-range entanglement associated with the N\'{e}el-ordered ground state becomes relevant\cite{LaflorencieWLR2006}.
How to improve the TTN structure by taking account of the entanglement between distant blocks without the increase of the computational cost is an essential problem for the E-tSDRG, which is left for future studies.
From the technical viewpoint, a parallelization of the process of the update of the list of entanglement entropy $\mathcal{S}_{r,r'}$ in E-tSDRG or energy gap $\Delta_{r,r'}$ in H-tSDRG may be rewarding.
In addition, it is effective to implement the variational optimization of the tensors by up-and-down sweeps of the tensor network with integrating unitary disentanglers\cite{TagliacozzoEV2009,GoldsboroughE2017}.

While our focus in this paper was on the benchmark of the tSDRG algorithms within the finite-cluster level, it is another important problem to explore how the nature and ability of the tSDRGs change as the system size is asymptotically larger.
For instance, it is an intriguing question what types of fixed point can be realized in the tSDRGs.
Since the H-tSDRG is a straightforward extension of the perturbative SDRG\cite{MaDH1979,DasguptaM1980}, it is expected that the H-tSDRG and also the E-tSDRG are capable of realizing the random-singlet fixed point\cite{Fisher1994}, which is established to be asymptotically exact in the framework of the perturbative SDRG.
On the other hand,  the present results could not give a clear answer for whether or not the E-tSDRG algorithm can properly describe the fixed points of magnetically-ordered states\cite{LaflorencieWLR2006} and 
the frustrated random-singlet state
\cite{WatanabeKNS2014,ShimokawaWK2015,KawamuraWS2014,UematsuK2017,UematsuK2018,KawamuraU2019,UematsuK2019,UematsuHK2020}.
It is also an essential issue to determine the asymptotic scaling form of key quantities, including the distribution of the entanglement entropy $\{ \mathcal{S}_{r,r'} \}$, the energy gap $\{ \Delta_{r,r'} \}$, and the block coordination number $\{ D_{r,r'} \}$, as functions of the iteration number in the tSDRG calculations.
We hope that the present work also stimulates further studies for understanding such exotic states induced by randomness.

\begin{acknowledgments}
This work was supported by JSPS KAKENHI Grant Numbers JP15K05198, JP17H02931, and JP19K03664.

\end{acknowledgments}



\begin{thebibliography}{38}
\bibitem [{deg()}]{degen_eigenstates}%
  \BibitemOpen
  \bibinfo {note} {Two comments are in order. First, at the early stage of the
  calculation, the dimension of the Hilbert space of the block pair $(R,R')$ to
  be normalized can be smaller than or equal to $\chi$. When this is the case,
  we do not truncate the Hilbert space of the block pair and keep it for the
  new block. Second, in the construction of the bases of the new block, we keep
  or discard all the degenerate eigenstates of $\mathcal{H}^{\rm P}_{R,R'}$ or
  $\tilde{\rho}_{R,R'}$ belonging to the same multiplet in order to preserve
  the SU(2) symmetry of the ground state. Therefore, the dimension of the
  Hilbert space of the new block is equal to or less than $\chi$, in
  practice.}\BibitemShut {Stop}%
\end{thebibliography}


%

\end{document}